\DocumentMetadata{}
\documentclass[sigconf]{acmart}
\usepackage{soul}
\usepackage[inline]{enumitem}
\usepackage{subcaption}

\copyrightyear{2024}
\acmYear{2024}
\setcopyright{rightsretained}
\acmConference[MOBILEHCI Adjunct '24]{26th International Conference on Mobile Human-Computer Interaction}{September 30-October 3, 2024}{Melbourne, VIC, Australia}
\acmBooktitle{26th International Conference on Mobile Human-Computer Interaction (MOBILEHCI Adjunct '24), September 30-October 3, 2024, Melbourne, VIC, Australia}
\acmDOI{10.1145/3640471.3680232}
\acmISBN{979-8-4007-0506-9/24/09} 

\begin{document}

\title{DriveStats: a Mobile Platform to Frame Effective Sustainable Driving Displays}

\author{Song Mi Lee-Kan}
\authornote{Author is currently at University of Michigan.}
\email{songmil@umich.edu}
\orcid{0000-0002-2829-8975} 
\affiliation{%
  \institution{Toyota Research Institute}
  \streetaddress{4400 El Camino Real}
  \city{Los Altos}
  \state{California}
  \country{USA}
  \postcode{94022}
}
\author{Alexandre Filipowicz}
\email{alex.filipowicz@tri.global}
\orcid{0000-0002-1311-386X} 
\affiliation{%
  \institution{Toyota Research Institute}
  \streetaddress{4400 El Camino Real}
  \city{Los Altos}
  \state{California}
  \country{USA}
  \postcode{94022}
}
\author{Nayeli Bravo}
\email{nayeli.bravo@tri.global}
\orcid{0000-0001-9238-9831} 
\affiliation{%
  \institution{Toyota Research Institute}
  \streetaddress{4400 El Camino Real}
  \city{Los Altos}
  \state{California}
  \country{USA}
  \postcode{94022}
}
\author{Candice L. Hogan}
\email{candice.hogan@tri.global}
\orcid{0000-0002-3240-2560} 
\affiliation{%
  \institution{Toyota Research Institute}
  \streetaddress{4400 El Camino Real}
  \city{Los Altos}
  \state{California}
  \country{USA}
  \postcode{94022}
}
\author{David A. Shamma}
\email{ayman.shamma@tri.global}
\orcid{0000-0003-2399-9374} 
\affiliation{%
  \institution{Toyota Research Institute}
  \streetaddress{4400 El Camino Real}
  \city{Los Altos}
  \state{California}
  \country{USA}
  \postcode{94022}
}

\renewcommand{\shortauthors}{Lee-Kan, et al.}
\newcommand{\mc}[0]{\hl{[missing cite]}}
\newcommand{\coo}[0]{CO\textsubscript{2}}

\begin{abstract}
    Phone applications to track vehicle information have become more common place, providing insights into fuel consumption, vehicle status, and sustainable driving behaviors.
    However, to test what resonates with drivers without deep vehicle integration requires a proper research instrument.
    We built DriveStats: a reusable library (and encompassing an mobile app) to monitor driving trips and display related information.
    By providing estimated cost/emission reductions in a goal directed framework, we demonstrate how information utility can increase over the course of a 10 day diary study with a group of North American participants.
    Participants were initially interested in monetary savings reported increased utility for emissions-related information with increased app usage and resulted in self-reported sustainable behavior change.
    The DriveStats package can be used as a research probe for a plurality of mobility studies (driving, cycling, walking, etc.) for supporting mobile transportation research.
\end{abstract}

\ccsdesc[500]{Social and professional topics~Sustainability}
\ccsdesc[500]{Human-centered computing~Field studies}
\ccsdesc[500]{Human-centered computing~User studies}
\ccsdesc[300]{Human-centered computing~Empirical studies in ubiquitous and mobile computing}

\keywords{sustainability, information, utility, carbon, cost, automotive, mobile, app, platform, survey, diary study}


\maketitle


\section{Introduction}
Personal transportation is the largest contributor to people's overall carbon footprint in many high-income countries~\cite{actnow}. 
Significant opportunities exist to create mobile apps for everyday drivers to reduce emissions through sustainable driving practices, eco-driving, and post-purchase behaviors~\cite{SIVAK201296}. 
Active eco-driving---mindful management of acceleration, braking, and cornering---immediately improves energy efficiency and lowers greenhouse gas (GHG) emissions~\cite{SIVAK201296}. These behavioral interventions (i.e., coaching) can happen in cabin or in a mobile app~\cite{AFWAHLBERG2007333,10.1115/1.3622753}.
Mobile apps make it easier to track and display various driving performance data and metrics, providing drivers opportunities to adjust behavior for eco-driving and safety~\cite{elvik2014rewarding,ellison2015driver,bergasa2014drivesafe}. 
However, disengagement can occur through pre-existing beliefs, unwelcome actions, and the information's lack of personal relevance~\cite{narayan2011role,sweeny2010information,golman2017information}. 
To understand how drivers perceive the personal utility of sustainability information, one needs a research probe to reflect daily driving habits and promote eco-behaviors.


In this study, we aim to test a goal-directed information framework~\cite{sharot2020people} focusing on major eco-driving incentives: fuel cost savings and \coo\ emissions reduction. 
We developed a reusable library and iOS mobile app called \textit{DriveStats} that we deployed in a 10-day diary study, where 27 drivers monitored potential fuel costs and \coo\ savings based on their vehicle usage. The app featured personalized dashboards for both monetary and carbon information, and suggested saving goals over 3-day sliding windows. 
Drivers' engagement with these displays were captured through online pre-and post-surveys and diary entries as well as in-app dwell time logs.
Overall we find that although people tend to have more utility for information about costs, time with the app and goal-directed framings help increase utility towards emission information, making people more willing to adopt eco-friendly driving behaviors.
Overall, we find that goal-directed framing amplifies information utility for eco-driving and promotes interest in \coo\ savings.
Further, we demonstrate and release the DriveStats iOS package as an open-source platform 
for creating in mobility-based research probes.
We find in a goal-directed, eco-driving, and \coo\ savings is amplified and information utility is increased.
Further we assert one must test such information utility in real-world settings and we demonstrate a research probe as a proxy-application for doing so.


\section{Related Work}
While many commercial automakers make companion apps for their gas, hybrid, and electric vehicles. Without vehicle intergration, research in sustainable driving is accomplished in simulators~\cite{van2001prototype,vaezipour2018simulator,10.1145/3543174.3545256} where the effects can be short lived~\cite{af2007long,degraeuwe2013corrigendum}.
Personal mobile health apps have had their successes and failures as they are tied into personal habits and social networks~\cite{10.1145/2800835.2800943,10.1145/3490632.3490671,10.1145/2628363.2645670}.
To seek a similar embodiment for \coo\ vehicular research, we aim to follow a similar pattern: a mobile app that accomponies individual drivers and reports money and sustainability information from personal driving.

\subsection{Drivers' Interest in Money versus Carbon Information}
Research and public polls consistently find that drivers are motivated by the monetary and environmental benefits of eco-driving and electric vehicles~\cite{10.1145/3544548.3581301,Peters2018-rv,31120589,Dogan2014-cp,stillwater2013drivers,Kramer2023-yv}. 
Lee et al.~\cite{10.1145/3544548.3581301} found that financial costs have a greater impact on decisions to purchase and charge plug-in hybrid vehicles, even for those identifying as environmentalists. However, although people express interest information about cost and \coo\ savings, this information does not always influence behavior. Dogan et al.~\cite{Dogan2014-cp} reported no significant difference between the impact of environmental and financial incentives on the intention to adopt eco-driving.
Cost alone shows little to influence people's driving behavior~\cite{stillwater2013drivers,kramer2023environmental}.
One caveat in previous findings is the absence of a systematic approach to understand why different types of information receive varying levels of attention, often leading to inconsistent or weak intervention outcomes. 
For example, \coo\ may not often enter daily considerations, resulting in less attention due to a lack of personal relevance. 

\subsection{Instrumental, Hedonic, and Cognitive Utilities of Information}
Sharot, Sunstein, and colleagues~\cite{sharot2020people,kelly2021individual} developed an integrative framework to assess individuals' motivations for seeking and avoiding information. 
This framework suggests that people are drawn to information they expect to provide positive ``utility'' across three key dimensions:
\begin{enumerate*}
    \item \textbf{Instrumental utility}: the practical usefulness of information in guiding actions toward external rewards or away from losses. 
    \item \textbf{Hedonic utility}: the emotional impact of information, sought after for positive emotions (e.g., joy from good news) and avoided when likely to induce negative feelings (e.g., distress from bad news).
    \item \textbf{Cognitive utility}: how well information serves one’s mental models or important concepts. People are drawn to information related to topics they consider frequently.
\end{enumerate*}  
The overall utility of information is calculated as the weighted sum of these three utilities, with weights varying by individual~\cite{sharot2020people,kelly2021individual}.
Some individuals may prioritize information that offers practical benefits (instrumental utility), while others may value information that improves their emotional state (hedonic utility) or aligns with their interests (cognitive utility). People engage with personal information if its overall utility is positive, avoid it if negative, and remain indifferent if neutral.

\subsection{Goal Framing May Enhance Information Utility}
While Sharot and Sunstein's framework~\cite{sharot2020people} offers deeper insight into the appeal of information, it does not provide strategies to enhance engagement with information perceived to have low utility---a significant challenge for sustainability initiatives. In the U.S., surveys indicate that environmental issues rank low in personal or national priorities and the public often views sustainability information (e.g., climate change, \coo\ emissions) as having low utility~\cite{Tyson2023-lx,Poushter2021-ss}. 
A promising strategy to boost information utility is goal-directed interventions, which concretize the potential outcomes of actions~\cite{lindenberg2007normative,noordzij2021meta}. 
A goal consists of a clear objective (e.g., reducing carbon emissions by 10\%) and can be framed positively or negatively (e.g., gaining health benefits or avoiding health risks~\cite{Tuk2021-im,noordzij2021meta}). Goals are most effective for behavior change when they are deemed desirable, important, and feasible~\cite{Levin1998-nl,bergquist2023field,Ploll2023-hl,Lindenberg2013-pw,Krishnamurthy2001-zx,haugtvedt2018goal, bagozzi2003effortful,pinder2018digital,locke2002building}, even among those who previously felt unconnected to the topic of the goal. 
As such, testing an information framework via a mobile application should include tests of goal-framing.

\section{DriveStats Mobile Platform and Application}
\begin{figure*}
  \centering
  \begin{subfigure}[t]{.4\columnwidth}
    \includegraphics[width=\columnwidth]{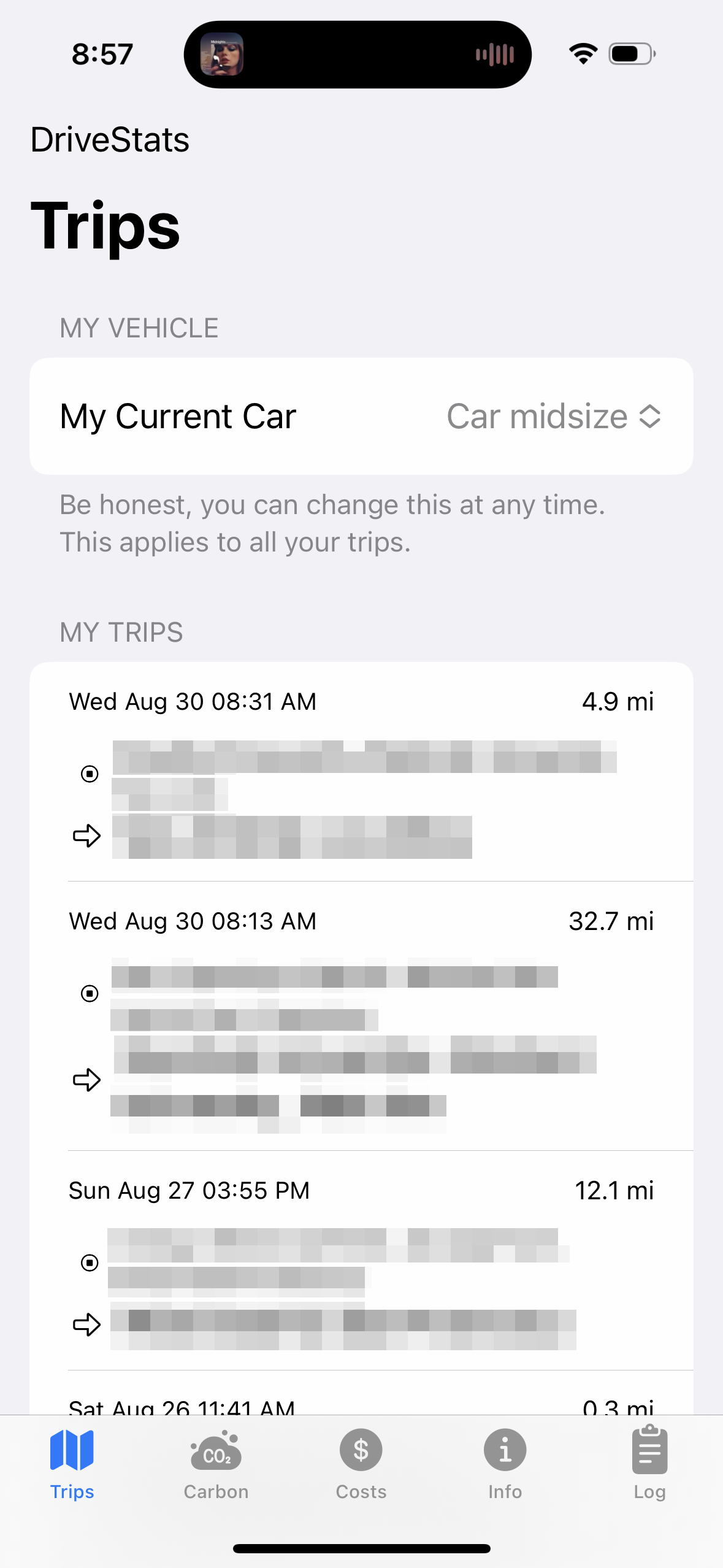}
    \caption{\label{fig:trips}}
  \end{subfigure}
  \hspace{.5pc}
  \begin{subfigure}[t]{.4\columnwidth}
    \includegraphics[width=\columnwidth]{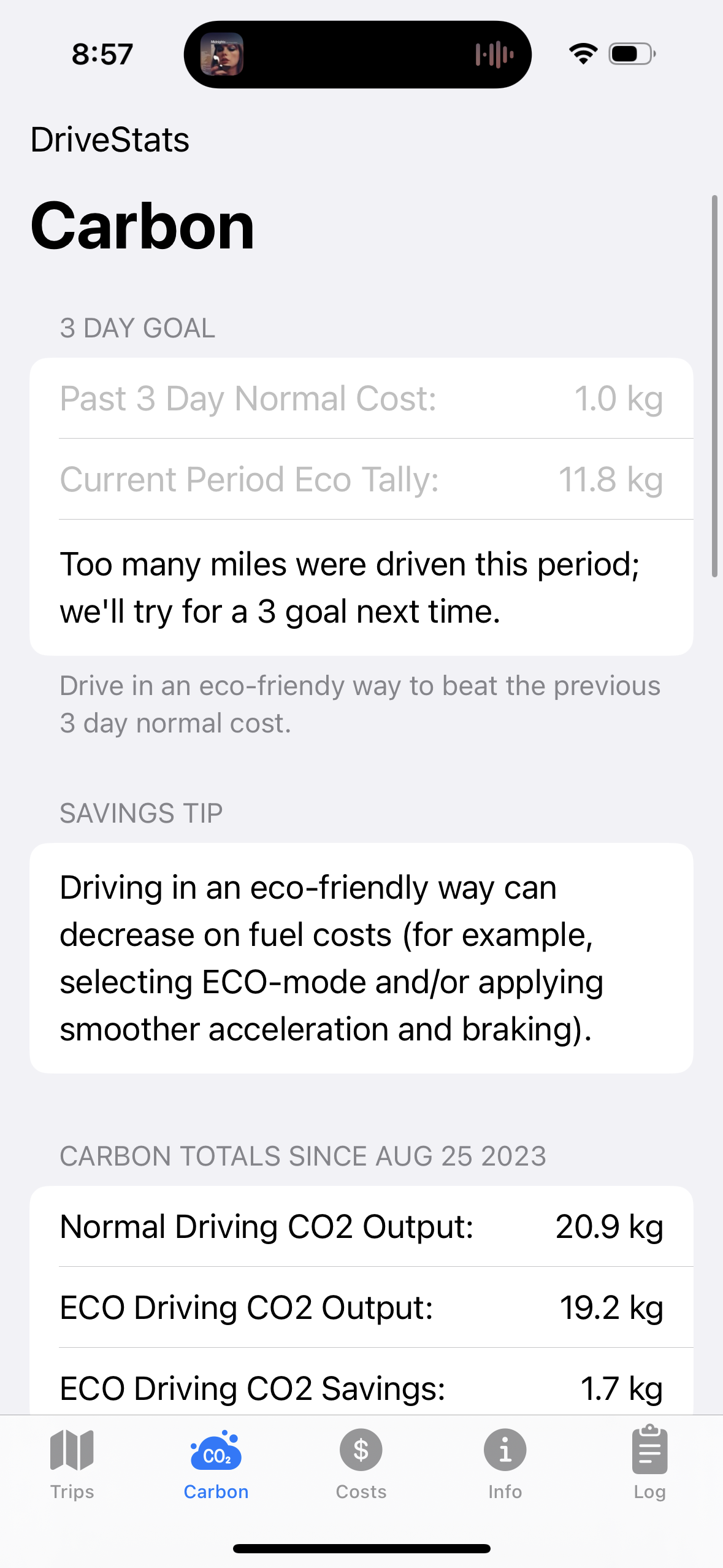}
    \caption{\label{fig:tabc}}
  \end{subfigure}
  \hspace{.5pc}
  \begin{subfigure}[t]{.4\columnwidth}
    \includegraphics[width=\textwidth]{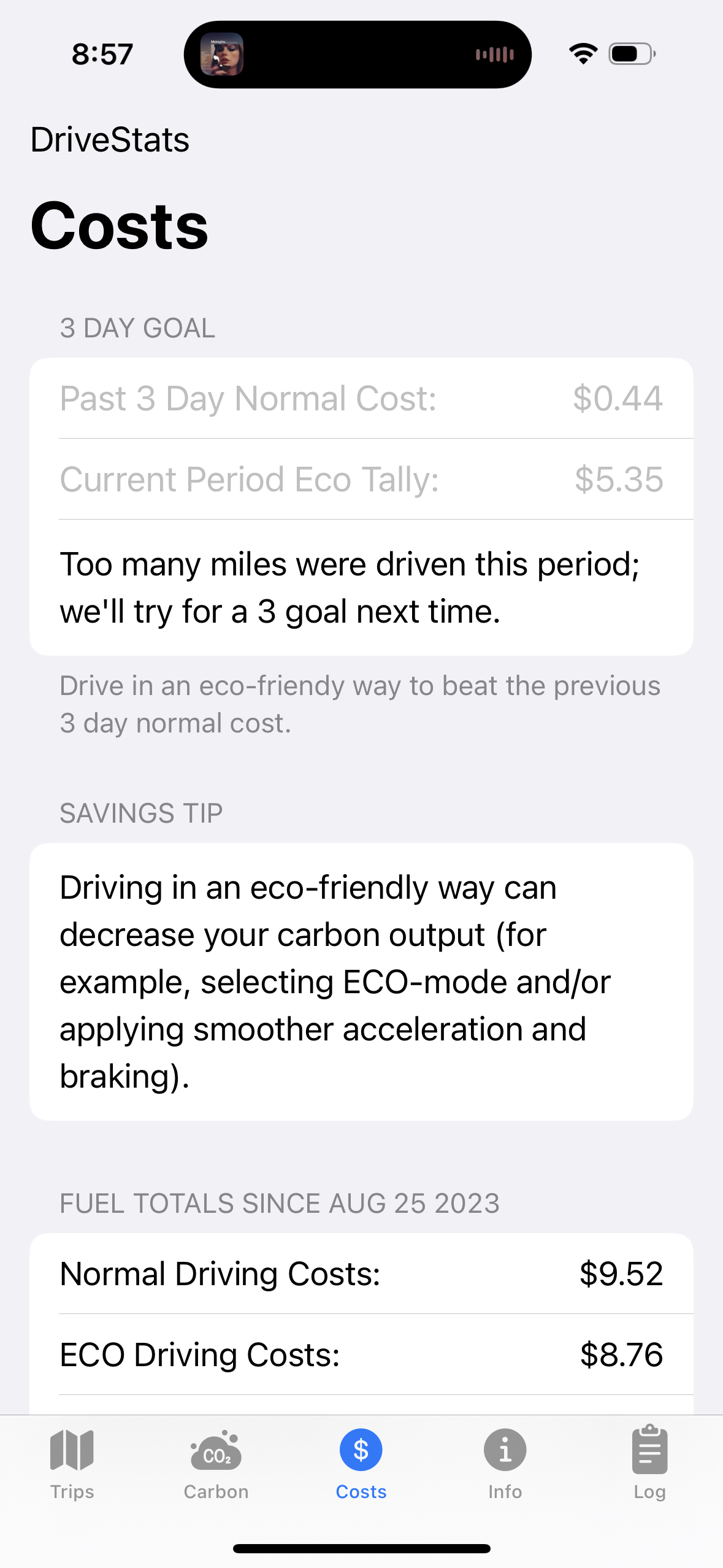}
    \caption{\label{fig:tabd}}
  \end{subfigure}
  \caption{\label{fig:tabh} The main tabs of the DriveStats app for the diary study.\ (\subref{fig:trips}) Trips tab shows to and from points with times and estimated distances.\ (\subref{fig:tabc}) Carbon tab displays a spent ``goal'' with totals.\ (\subref{fig:tabd}) Costs tab displays a spent ``goal'' with
      totals.}
  \Description{Three tabs from the app. The first showing trips (redacted for privacy). The second showing Carbon and the third showing Cost.  The latter two show the 3-day goal at the top, a small driving tip for eco, and then the full counts below.}
\end{figure*}

\begin{figure}
  \centering
  \includegraphics[width=.9\columnwidth]{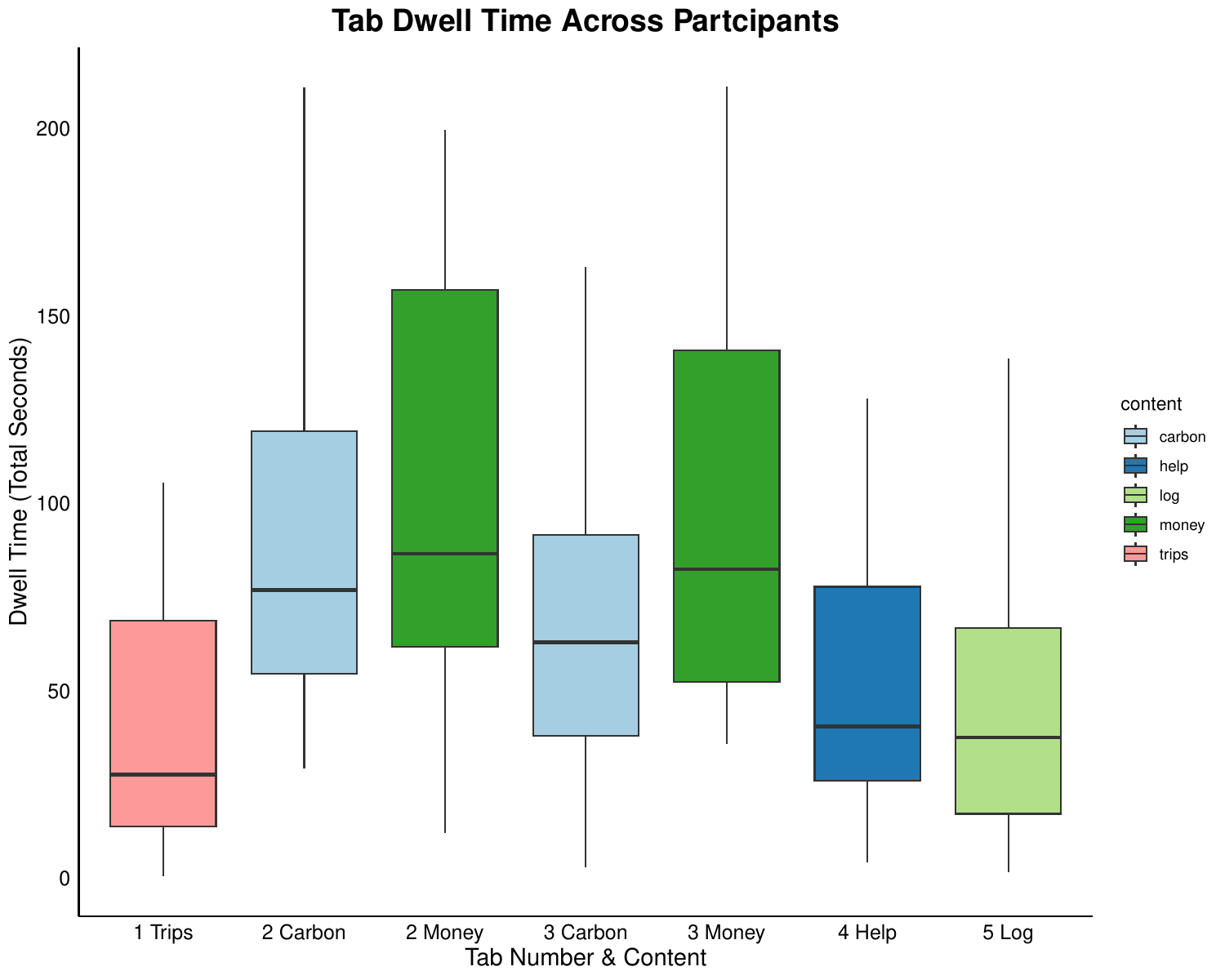}
  \caption{\label{fig:lutes} Box plots of the dwell times across all the tabs. Participants spent significantly more time on the money tab over the carbon tab ($p=0.0117$). Tabs Carbon and Money were randomized across the population to mitigate order bias.}
  \Description{A box plot showing the dwell times on each tab. The IQR for money tabs exceeds the others.}
\end{figure}

In this study, we explore how drivers' perceptions of utility regarding monetary and \coo\ information evolve over time and assess how these perceptions influence their motivation for eco-driving within their personal driving contexts.
Additionally, we evaluate whether integrating this information with specific goals enhances its impact.
For this, we created an iOS package and app called \textit{DriveStats} (Figure~\ref{fig:tabh}), which provides users with summaries of fuel costs and \coo\ from their vehicle trips. DriveStats aims to promote more sustainable driving choices by offering personalized insights into fuel costs, emissions, and potential savings of these two.
Modern smartphones facilitate automatic logging of trip data, which is essential for accurately calculating fuel consumption and emissions. While direct vehicle interfacing via APIs like SmartCar~\cite{smartcar} or Bluetooth OBD-II loggers~\cite{odb2} is possible, this approach can introduces complexities, such as the need for vehicle synchronization and privacy concerns.
We applied our previous architecture of an application for vehicle interventions~\cite{10.1145/3490100.3516451} and created a custom front-end tailored to our study's needs.
This setup allowed us to focus on behavioral insights without technical complications associated with direct vehicle interfacing.

\subsection{Identifying Car Trips}
To identify car trips, we used Apple's CoreLocation library to capture significant location changes using GPS coordinates and departure-arrival timestamps.
We used the CoreMotion API to distinguish automotive activities from non-fuel-based motions such as walking and biking, and Mapkit to estimate the most probable routes taken between locations.
We opted against live GPS tracking to avoid user privacy issues and excessive battery drain, ensuring the chosen APIs provided battery-efficient and sufficiently accurate data collection.
Participants were consistently reminded that data shown in DriveStats are estimates based on probabilities and national averages, which might differ from actual metrics.

\subsection{App Displays}
The DriveStats app was built and distributed exclusively via TestFlight, Apple's beta testing platform, for research purposes. 
The app was built using Apple's basic SwiftUI framework, focusing on evaluating information needs rather than design aesthetics.
The app was modeled from another research platform example~\cite{10.1145/3490100.3516451}.
DriveStats features five tabs. The first \textit{Trips} tab (Figure~\ref{fig:trips}) lists all a user's vehicle trips, with options to delete entries for privacy or irrelevance (e.g., taxi rides). In this tab, users can specify or change their vehicle type, which is used to calculate the fuel costs and carbon emissions of each trip. 
The next two tabs, randomly ordered at the first launch to prevent display order bias, summarize carbon output (\textit{Carbon} tab, Figure~\ref{fig:tabc}) and fuel costs (\textit{Costs} tab, Figure~\ref{fig:tabd}).
Each tab displays a running total of estimated emissions and expenses, alongside potential reductions or savings if eco-mode was enabled or if the the driver self-reported eco-friendly by following tips in the app.
These tabs also detail trip-by-trip data.
After three days of logging, a snapshot of the previous three days' fuel costs and carbon emissions is displayed at the top of each tab as a dynamic goal for users to achieve.
If a user drives excessively, resulting in high emissions or costs, a notification suggests aiming for improvement in the next period.
The fourth \textit{Info} tab provides details about the app, eco-driving tips, and instructions on how to delete trips as needed.
The final \textit{Research Log} tab records CSV data, including Unix timestamps and tags indicating user interactions (e.g., tab clicks or changes in the app's foreground/background status).
At the study's conclusion, participants submitted this log through the app.
We analyzed these logs to measure dwell times, assessing how long participants engaged with different app tabs.

\subsection{Computing Carbon and Fuel Costs}
Fuel costs and carbon emissions are computed based on a list of generated trips. The process involves initial calculations of the cost and emissions for each trip, followed by computations of potential savings with eco-driving practices. The following methods ensure that our estimates of potential fuel and carbon savings are both realistic and tailored to the specific driving conditions and vehicle types of the participants.
\textbf{Fuel cost estimation}: Fuel costs per mile are estimated using statistics from the U.S. Energy Information Administration~\cite{epa:gasdiesel}, with a reference price of \$3.85/gallon  as the national average at the time of this study.
\textbf{Carbon Emission Calculation}: We utilized CarGHG~\cite{carghg} to calculate fuel costs and \coo\ emissions per mile for various vehicle types. CarGHG bases its calculations on an annual driving average of 13,500 miles, which was also adopted by our app.
The calculations were expanded to include several vehicle types and powertrains using the EPA Green Vehicle Guide~\cite{doe:fueleconomy}, covering Internal Combustion Engine (ICE) and Hybrid Electric Vehicles (HEV). Battery Electric (BEV), Plug-in Hybrids (PHEV), and fuel cell vehicles were excluded from this study.
\textbf {Vehicle Selection in DriveStats}: The app enables users to identify their current vehicle from various categories including small, midsized, and large cars, SUVs, and minivans, all available with ICE and HEV options, as well as trucks, station wagons, and sports cars.
\textbf {Computing Eco-Savings}: 
We computed potential eco-savings based on previous research finding~\cite{ivanov2019study} that eco-driving modes can reduce fuel consumption by 7\% to 17.5\% in city driving, and 3.9\% on highways
We computed three savings tiers: trips under 5 miles were labeled as non-highway and displayed 17.5\% in potential savings, trips over 15 miles were labeled as highway and given 3.9\% savings, and trips 5--15 mile were given savings on a linear scale, ranging 17.5\% to 3.9\% as mileage increased.

\subsection{Privacy Considerations}
Any research application that stores trips as GPS data requires extra careful consideration due to privacy concerns. In this study, the primary focus was on diary recordings to understand how participants used the app and how the displayed information influenced their thoughts and feelings. Consequently, the app operated entirely client-side, storing no GPS data on remote servers. Participants could delete any trip or location directly within the app by swiping left on the entry, ensuring control over their data. They were informed about the app's client-side nature and their ability to remove data before installation.
The application maintained a local log of the following user interactions for analysis:
\begin{enumerate*} 
\item Application to foreground
\item Application to background
\item Tab focus: Trip, Carbon, Cost, Info, or Log.
\end{enumerate*} No other data was captured. At the study's conclusion, participants manually copied and pasted this log to the researchers, reinforcing the transparency of the data handling process and ensuring that no GPS data was involuntarily extracted.


\section{App Diary Study}

We recruited 36 participants from the DScout usability testing platform~\cite{dscout} for a 10-day diary study, offering a 200 USD incentive for completion. 
Participants varied across demographics, with 23 identifying as women, one as non-binary, and 12 as men, aged between 20 to 71 years. 
They were selected to represent a balanced mix of age, gender, and driving areas (urban, suburban, or rural); however all drivers resided in the USA which presents a limitation on the sample.
All participants were daily drivers: 69.4\% reported sometimes driving in urban areas, 72.2\% in suburban areas, and 16.7\% in rural areas. Each participant owned and regularly operated either an HEV or ICE vehicle, while drivers of PHEV and BEV were excluded from the study due to the different ways fuel costs apply to these vehicles. Educational backgrounds included 13 college graduates and 9 postgraduates, while income levels were mostly above \$50,000 annually.
All participants carried iPhones with iOS 16+ versions.

Out of the 36 individuals who agreed to participate in the study, three dropped out for unspecified reasons, and two were dismissed with partial compensation owing to technical issues with the app. Two other participants did not complete the final exit survey. Several others (P8, P19, P22, P27, P32) encountered glitches but were guided to complete the tasks to the best of their abilities. Ultimately, 27 participants successfully completed all required activities. 

Participants were guided through the installation of the DriveStats app, setting necessary permissions, and learning usage guidelines, including not interacting with the app while driving.
They were to use the app after taking a trip in their car and no data would update in the app until several minutes after driving was no longer detected.
They engaged with DriveStats over 10 days in the summer of 2023, completing a series of activities as part of a diary study.

\begin{enumerate*}
  \item\label{li:presurvey} \textbf{Pre-launch survey}: Upon recruitment,
  participants were asked to estimate the utility they expected to get from two
  types of information:
  \begin{enumerate*}[label={(\alph*)}]
  \item potential savings on fuel costs and
  \item possible reductions in carbon emissions while driving.
  \end{enumerate*}

  We employed the existing survey instrument~\citep{kelly2021individual} to assess in interests in Instrumental, Hedonic, and Cognitive utilities.
These were asked on a $-3$ to $+3$ Likert scale.
  Participants also reported their driving styles and motivations to use self-tracking tools, which may influence their engagement with the app irrespective of information utilities. These propensities were gauged using the pre-validated Impulsive-Driving Scale~\cite{perez2015impulsive} and a set of self-tracking motivation questions~\cite{gimpel2013quantifying}. 
  

\item\label{li:diary} \textbf{Diary entries}: Participants were required to submit a minimum of one diary entry every two days, documenting what prompted them to check DriveStats, alongside a screenshot and a selfie-style video explaining their interest in the specific app display. Participants who were unable to install the app (via TestFlight) and grant location permissions were disqualified.

\item \textbf{Midpoint check-in}: A brief survey assessed the app's usefulness and any changes in driving behavior or perceptions, and planned continue use of the app. 

\item \textbf{Exit survey}: Participants needed to submit at least two additional diary entries. The exit survey revisited the utility measures from the pre-launch survey.
  
\item \textbf{Application log}: App usage data (e.g., app open events and dwell time) were stored locally on participants' phones. Participants were required to copy-paste-upload this running log into a dairy entry. Following the study, the app self-deactivated and instructions were provided to delete the app. 
\end{enumerate*}


\section{Results}

\subsection{Pre-Post Survey Analysis}

To quantitatively evaluate the impact of the DriveStats app on perceived information utilities, we conducted paired t-tests comparing pre-launch and exit survey responses. 
 The results revealed a significant increase in how often participants thought about fuel cost savings after using the app (post-pre mean = 0.58, p = 0.02, 95\% CI = [0.10, 1.06]), indicating an increased cognitive utility of this information. Other types of utilities of monetary information, such as usefulness (instrumental) or emotions (hedonic) showed no significant changes. Notably, there was a significant increase in participants' preference \textit{not} to know about potential carbon emissions reductions after the intervention (post-pre mean = 0.94, p < 0.001, 95\% CI = [0.34, 1.53]), while other utilities of \coo\ information did not change much. This may reflect growing discomfort associated with seeing carbon output data, possibly exacerbating feelings of obligation or pressure to improve, which contrasts with the anticipated desensitization through repeated exposure.

\subsection{Dwell Time}\label{sec:dwell-time}
In-app activity logs tracked events such as tab openings (with the \textit{Trips} tab as the default) and instances of the app moving to the background or foreground. 
For this experiment, data was separated into tabs to encourage user engagement based on perceived utility and to facilitate dwell time analysis.
Analysis revealed that participants spent more time on the \textit{Costs} tab compared to the \textit{Carbon} tab (Median-money: 86.7s; Median-carbon: 73.4s; Wilcox $W = 121$, $p=0.0117$; Fig.~\ref{fig:lutes}). A slight display order effect was observed despite attempts to mitigate it by randomizing the order of the \textit{Money} and \textit{Carbon} tabs. The second tab displayed during a session consistently had a longer dwell time than the third (Median-second: 80.8s; Median-third: 73.4s; Wilcox $W = 349$, $p=0.04789$), irrespective of its content.

\subsection{Diary Entries}
Qualitative feedback from diary entries uncovered nuances that were not highlighted as significant in the quantitative analyses.
DriveStats tracks user trips and displays monetary and carbon data summarizing overall (from the start of the experiment), 3-day, and trip-by-trip totals.
Beyond UI considerations, the app details costs and potential savings from eco-driving, providing tips for more the use of eco drive mode and active eco-driving practices.


\subsubsection{Eco-Driving and Eco Mode.}
By the midpoint check-in, most participants recalled, and reported practicing, at least one in-app eco-suggestion.
However, participants noted a decline in actionable information as they continued using the app, attributing this to the static nature of the recommendations.
They expressed a desire for more dynamic and detailed information about their driving habits.
Enhancements such as tracking hard braking, providing personalized driving recommendations, and offering comparisons with other drivers were suggested to increase the app's value.
Some drivers, (P13 and P17), already considered themselves safe, conscious, and smooth drivers, before using the app.
Of the 14 participants discussing eco mode, 11 viewed it favorably.
P21 discovered their car had an eco mode and started using it, while P5 and P23 reported activating eco mode became a habit midway through the study.
Conversely, P2 and P15 felt the minor savings didn't justify its use.
P24 preferred driving in ``normal mode'' for a more enjoyable driving experience and P7 felt eco-mode would damage their vehicle.
These comments illustrate a range of perceptions and experiences with eco mode, from skepticism about its benefits to concerns over its impact on vehicle longevity.

\subsubsection{Money}
Participants showed a stronger interest in saving money over carbon. 
At the midpoint, 18 out of 27 drivers recalled cost savings with eco mode in the greatest detail without opening the app and monetary savings were brought up by P5, P7, P9, P20, P23, P26, P29, and P31.
The app provided various perspectives on savings, emphasizing an accumulating effect (P9, P23, P26, P29, P31).
Similarly, P26 and P29 started to reconsider their driving habits altogether, with P26 saying, ``Understanding trip cost changes my decision, not just how I drive, but also if I would drive [at all].''

\subsubsection{Carbon}
Many participants found the \coo\ tally increasingly interesting, though some found it confusing. 
Being curious about carbon emissions was one of the most frequently mentioned triggers for checking the app (12 mentions), along with ``after a long drive or road trip'' (15 mentions) and ``before bed'' (13 mentions). 
Also, 13 out of 27 vividly recalled estimated \coo\ per trip without opening the app by the study's midpoint; this recall was also highlighted in the diary study as participants stated \coo\ was more on their mind (i.e.\ exhibited a higher cognitive information utility).
P20 appreciated having a personal \coo\ gauge.
P14 stated \coo\ savings were small but ``significant,'' further wondering about ``a lot of cars' incorporated savings.'' 
P3 expressed a desire for accessible, relatable information that connects daily activities to global carbon reduction efforts.
Not knowing what a kilogram of \coo\ is led to further inquiries
P1 stated ``I don't know. Did I save a tree?\ldots Am I doing a good job?'' 
Collectively, participants (like P1, P12, and P33) demonstrated a desire for more education on carbon impact~\cite{mohanty2023save}.
One participant (P28) stated \coo\ was ``not something I'm worried about.''

\subsubsection{Goals}
The app set goals by displaying a running total and a summary of the past three days, alongside overall totals. If a user exceeded their previous period’s sum, a message appeared: ``You drove more than last period, try again when the current period resets.''
The goal feature raised cognitive utility among some participants. 
P5 expressed interest in weekly goals, noting, ``It was interesting to see the potential savings over that time frame.''
This message was displayed at the top of the related Money or Carbon tab.
P31 found the goal motivating, appreciating the breakdown of each trip’s costs and potential savings. 
P11 pointed out statistics are ``kind of meaningless.''
However, they found the goal feature: ``one thing I actually found interesting\ldots was the last three days, like, fuel cost.''
P1 expressed dissatisfaction noting, ``\ldots I think, the app is telling me that I drove actually too much.''

\section{Discussion}
Modern connected cars carry in-cabin displays and mobile companion apps to let owners engage with information (and entertainment) beyond the traditional instrument cluster.
There is an opportunity to further use these displays to improve sustainability.  
While surveys can show money as a primary factor~\cite{10.1145/3544548.3581301}, we find that carbon savings in a goal-directed framework may increase information utility when put in the context of one's personal daily driving.
However, one must frame this discussion under the deployed scope of a 10 day diary study in the USA\@.

\subsection{Platform and Application}
To test information displays in the context of daily driving, we built DriveStats as a reusable research platform.
Utilizing native frameworks, the system determines vehicle trips without the need for deep car integration. 
Further, our design is anonymous and client-side, preserving privacy as there is no transmission of GPS coordinates back to the researchers and participants remain in control of their data.
Several participants assumed the app was vehicle-connected; we assured them this was not the case.
Currently, the platform is limited to iOS only but could be expanded further provided the API hooks required are present and engineering time can be spent.
For the study, the app set goals based on a 3-day non-sliding window.
A longer study would be required to establish history and a dynamic goal.
Further, we built the platform to be open-source and to monitor all mobility and trips including cycling, walking, and other detectable modes of transportation to support various research questions and experiments.
Additional development is also needed to incorporate other transportation types (commuter rails, water taxis, etc.) into the the framework via machine learning or applied heuristics, though during our diary study, these alternative transportation methods did not surface.

\subsection{Money Attracts, But Amount Matters}
Most participants were attracted to the fuel cost-saving aspect of eco-driving.
Cost savings were deemed useful in guiding actions such as planning more efficient routes and budgeting for future trips. Cost savings also remained in users' memory more frequently and often evoked positive emotions such as excitement, happiness, and pride.
However, participants frequently attributed these feelings not just to saving money but also to the simultaneous reduction in carbon emissions. 
This dual benefit suggests the environmental savings were viewed as a valuable additional benefit, providing a moral buffer for those primarily concerned with costs. 
This finding may contrast with previous research~\cite{Schwartz2015-bn,kramer2023environmental} that suggests monetary information undermines environmental concerns when it comes to sustainable behavior change.
Beyond comparing prices at the pump, participants wanted personalized cost savings.
Additionally, minor savings might be better reported as percentages~\cite{shu2022reducing,10.1145/3580585.3607176} to motivate positive driving habits.

\subsection{Carbon's Low Instrumental Utility, Promising Cognitive Utility, and Mixed Hedonic Utility}
\coo\ information utility was generally rated low as interpretation is arduous. 
Participants requested evaluative feedback such as ``good'' or ``bad,'' rather than raw numbers due to lack of reference.
Moreover, despite reductions in \coo\ emissions and fuel costs both deriving from reduced fuel consumption, participants often viewed them as separate issues, failing to see the correlation.
Participants' curiosity about carbon information increased through exposure to the app despite carbon's initial irrelevance.
Participants' perception of hedonic utility in \coo\ were particularly conflicting.
Some expressed positive feelings after learning about potential carbon savings, while others experienced negative emotions such as guilt or frustration. 
While negative emotions can lead to information avoidance~\cite{narayan2011role,sweeny2010information,golman2017information}, opportunities for effective interventions through better \coo\ communication~\cite{mohanty2023save,10.1145/3544999.3552530} and collective social interventions~\cite{10.1145/3543174.3545256,mohanty2023save} could thrive when linked to personal daily habits.

\section{Conclusions}
To study interventions relating to people's transportation behaviors, one must relate to their daily driving habits directly.  
To accomplish this, we present the DriveStats framework for rapid development of mobile applications relating to transportation, communing, or even physical activity (exercise or walking travel). 
Using this framework in an app-based diary study, we investigated drivers' perceived value for information about driving-related fuel cost and emission reductions, examining how these perceptions influence user engagement with the information and, ultimately, eco-driving behaviors. 
We found that interest levels vary across different forms of information utility (instrumental, hedonic, and cognitive) depending on the type of information presented. 
Additionally, we observed that people's perception of information utility perceptions can be increased with goal-framing. 
By unpacking what's `interesting' or `useful,' future interventions can be better tailored to meet user information needs. Finally, we aim to iterate on DriveStats as an open-source platform to enable future research endeavors for automotive research.

\begin{acks}
We would like to thank Fred Faust and Kalani Murakami for their help in developing the DriveStats platform. We also thank the anonymous reviewers for their constructive feedback.
\end{acks}

\bibliographystyle{ACM-Reference-Format}
\bibliography{main}

\end{document}